# Low-Frequency Current Fluctuations in *Graphene-like* Exfoliated Thin-Films of Topological Insulators

M. Z. Hossain,[1] S. L. Rumyantsev,[2,3] K. M. F. Shahil,[1] D. Teweldebrhan,[1] M. Shur[2] and A. A. Balandin[1,*]

[1]*Nano-Device Laboratory, Department of Electrical Engineering and Materials Science and Engineering Program, Bourns College of Engineering, University of California – Riverside, Riverside, California, 92521 USA*

[2]*Department of Electrical, Computer and Systems Engineering and Center for Integrated Electronics, Rensselaer Polytechnic Institute, Troy, New York, 12180 USA*

[3]*Ioffe Institute, Russian Academy of Sciences, St. Petersburg, 194021 Russia*

**Abstract**

We report on the low-frequency current fluctuations and electronic noise in thin-films made of $Bi_2Se_3$ topological insulators. The films were prepared via the "graphene-like" mechanical exfoliation and used as the current conducting channels in the four- and two-contact devices. Analysis of the resistance dependence on the film thickness indicates that the surface contribution to conductance is dominant in our samples. It was established that the current fluctuations have the noise spectrum close to the pure $1/f$ in the frequency range from 1 to 10 kHz ($f$ is the frequency). The relative noise amplitude $S_I/I^2$ for the examined $Bi_2Se_3$ films was increasing from ~$5\times10^{-8}$ to $5\times10^{-6}$ (1/Hz) as the resistance of the channels varied from ~$10^3$ to $10^5$ Ω. The obtained noise data is important for understanding electron transport through the surface and volume of topological insulators, and proposed applications of this class of materials.

Topological insulators constitute a new class of quantum materials, with insulating energy gaps in their bulk and gapless Dirac cone type surface states at the boundaries [1-10]. Topological insulator films exhibit quantum-Hall-like behavior in the absence of magnetic fields. This new class of materials was proposed for realization of the dissipation-less interconnects, low-power electronics and quantum computing using the surface states that are topologically protected against scattering by the time-reversal symmetry. They can also be used for magnetic memory where the write and read operations are achieved by purely electric means. Topological insulators have also shown promise for thermoelectric applications at low and room temperatures [11-13].

Bismuth selenide ($Bi_2Se_3$) and related materials such as $Bi_2Te_3$ and $Sb_2Te_3$ have been identified as three-dimensional (3D) topological insulators with a single Dirac cone at the surface [5]. Among 3D topological insulators, $Bi_2Se_3$ serves as a reference material for topological insulator behavior owing to its relatively simple band structure and large bulk band gap [5-7]. Compared to bulk, nanostructured $Bi_2Se_3$ offers an attractive alternative as topological insulator because of the high surface-to-volume ratio and tuning capability via thickness variation [14]. The rhombohedral crystal structure of $Bi_2Se_3$ belongs to the space group $D_{3d}^5(R\bar{3}m)$ with five atoms in the trigonal unit cell [15]. The structure is most simply visualized in terms of a layered structure with each layer referred as a quintuple layer (see Figure 1). Each quintuple layer with a thickness of ~1 nm consists of five atomic planes arranged in the sequence of Se(1)–Bi–Se(2)–Bi–Se(1). The coupling is strong within one quintuple layer but weak between any two quintuples, which are bonded by the van der Waals forces. The lattice parameters of the hexagonal unit cell are $a_H$=0.41384 nm and $c_H$=2.864 nm [15].

The layered structure of the crystal with van der Waals "gaps" allows one to mechanically exfoliate thin films with the thickness down to a single quintuple layer. Some of us have previously demonstrated "graphene-like" exfoliation of quintuples and few-quintuple layers of $Bi_2Te_3$ – materials with a similar crystal lattice [16-18]. A number of recent reports described the current–voltage characteristics of $Bi_2Se_3$ devices with the gate controlled chemical potential [19-21]. At the same time, despite its scientific and technological importance, a systematic study of the low-frequency current fluctuations and electronic noise in thin films and devices made of $Bi_2Se_3$ or other topological insulators still awaits its turn.

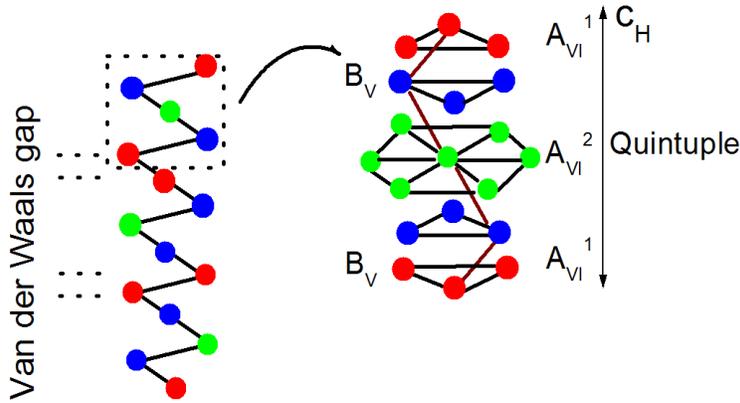

**Figure 1:** Crystal structure of the $B_V$-$A_{VI}$ compounds (i.e. $Bi_2Se_3$ or $Bi_2Te_3$) and a quintuple layer. Here $A_{VI}^2$ ($Te^2$, $Se^2$) atoms have the role of inversion centers.

Here, we report results of an experimental study of the low-frequency current fluctuations and $1/f$ electronic noise in thin films made of topological insulator material (here $f$ is frequency). Analysis of the resistance $R$ dependence on the film thickness $H$ indicates that the surface contribution to conductance is substantial in our samples. The knowledge of the $1/f$ noise spectra reveals information about the specifics of the charge current transport in topological insulators. From the applications point of view, understanding the low-frequency noise is essential for maintaining acceptable signal-to-noise ratio in topological insulator devices. We intentionally selected rather thick films, with the thickness ranging from ~50 to 170 nm, in order to avoid hybridization of the top and bottom electron surface states with the corresponding opening a gap on the surface. At the same time, the study of the thin films, rather than the large bulk conductors, allowed us to have channels with better surface uniformity and quality (the mechanically cleaved surfaces of this material are very clean) as well as to reduce the volume part of the electron conduction. The next section describes the results of the measurements followed by conclusions and methods.

**SAMPLE PREPARATION AND MEASUREMENTS**

We mechanically isolated thin films of $Bi_2Se_3$ following the procedure described by some of us for $Bi_2Te_3$ [16-18]. The exfoliated films had lateral dimensions of ~10 μm and the thickness larger than 50 nm. The films were identified using optical microscope and scanning electron microscope (SEM). The crystalinity and quality of the films were verified with micro-Raman spectroscopy [18]. We used the electron beam lithography system to define the four or two metal contacts for electrical measurements. Figure 2 shows the schematic of the fabricated devices. The distances between two adjacent contacts ranged from ~1 μm to ~3 μm while the topological insulator channel width varied from 1 μm to 16 μm. The thickness of the films was verified with the atomic force microscopy (AFM).

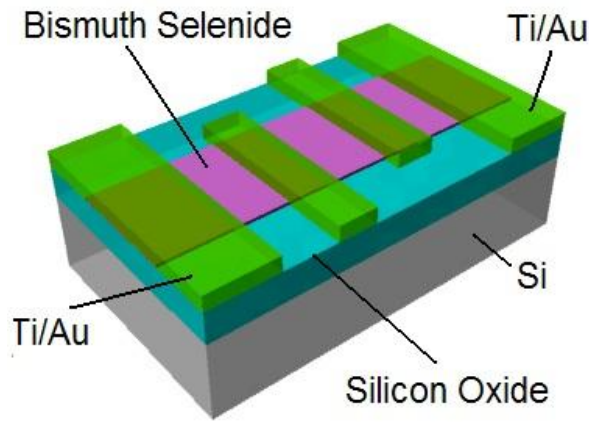

**Figure 2:** Schematic of the four-terminal devices fabricated with the thin films made of $Bi_2Se_3$ topological insulator material.

In Figure 3 (a-d) we present AFM, SEM and optical microscopy images of the variety of the examined devices. The current-voltage (I-V) characteristics of the devices were determined via four-terminal measurements to reduce the effects of the metal contacts. The tested devices revealed linear current-voltage characteristics for the low applied voltage with the resistance $R_{ti}$ from 500 Ω to $10^5$ Ω.

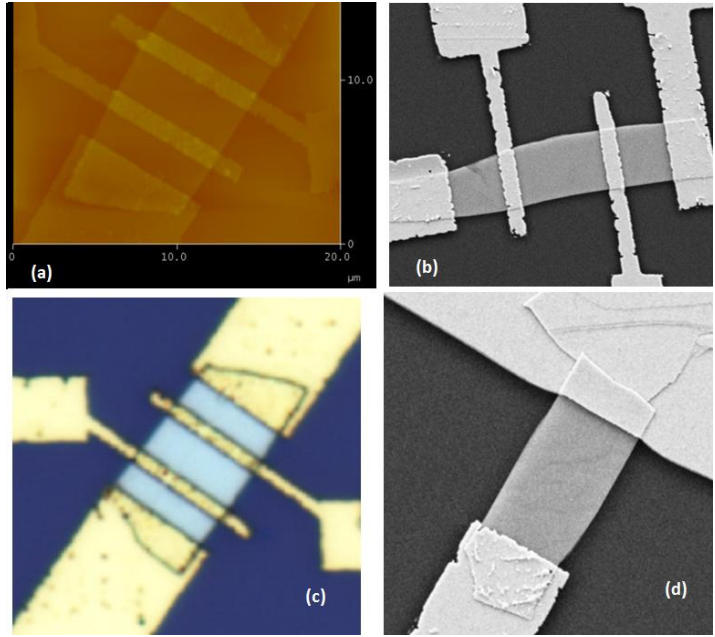

**Figure 3:** (a) AFM image of the four-terminal device. AFM inspection was used for determining the channel thickness. (b) SEM image of the four-terminal device structure with the topological insulator channel. (c) Optical image of the device under test. (d) SEM image of the two-terminal reference device with the $Bi_2Se_3$ channel.

Figure 4 shows an experimental setup used for electrical measurements and a schematic of the noise measurement procedure. The whole setup was placed inside the metal shielding box to reduce the effects of environmental noises and electromagnetic fields. The devices-under-test (DUT) were contacted by the micro-probes. The low-noise batteries and metal resistors were used for the noise measurements to provide the DC bias on the samples ranging from 5.8 mV to 40 mV. The signal from the load resistor $R_L$=1 kΩ - 10 kΩ was amplified by the low noise amplifier and analyzed by a spectrum analyzer (SR 770). The background noise of the amplifier was always 2-3 orders of magnitude smaller than the measured noise $S_v$ and was subtracted from the measured spectra.

For the data analysis it is important to understand where the main contribution to electronic conduction is coming from: volume or surface of the $Bi_2Se_3$ channel. For this reason, we measured the current voltage characteristics (I-V) of many devices with a different thickness $H$. Assuming that the resistivity $\rho$ of the material is approximately constant one can expect that in the case when electron transport is dominated by the volume contributions, the measured resistance $R$ should scale inversely proportional to the film thickness $H$. This is valid when the contact resistance $R_C$ does not dominate the measured resistance as was the case with our samples ($R_C<<R$). If the transport is dominated by the surface contributions it should be independent of the film thickness.

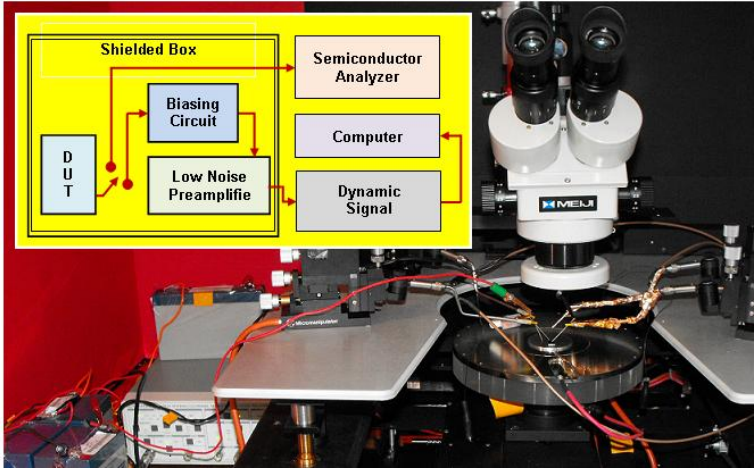

**Figure 4:** Experimental setup inside the metal shielding box and schematic of the measurement procedure. Note that the power is provided by the low-noise batteries.

In Figure 5 we show $(R/\rho)(W/L)$ as a function of $H$ ($L$ is the length of the channel and $W$ is the width of the channel). The absence of clear scaling with $H$ suggests that the surface contributions in our samples are significant, and, as such, are expected to dominate the low-frequency noise response. Two data points ($H$=81 nm and $H$=150 nm), which somewhat deviate from the general trend, correspond to the films with the lowest measured $R$. The latter suggests that in these two films the volume contribution was larger. It is difficult to assign a quantitative value to the surface and volume contributions because the assumption of constant $\rho$ is valid only approximately. Although the samples of different $H$ are exfoliated from the same bulk crystal, the Fermi level position may vary due to variations in the defect densities and surface condition as well as thickness non-uniformity over the channel length. The defects and thickness variations may lead to deviations from stoichiometry and, as a result, non-intentional doping and changes in $\rho$.

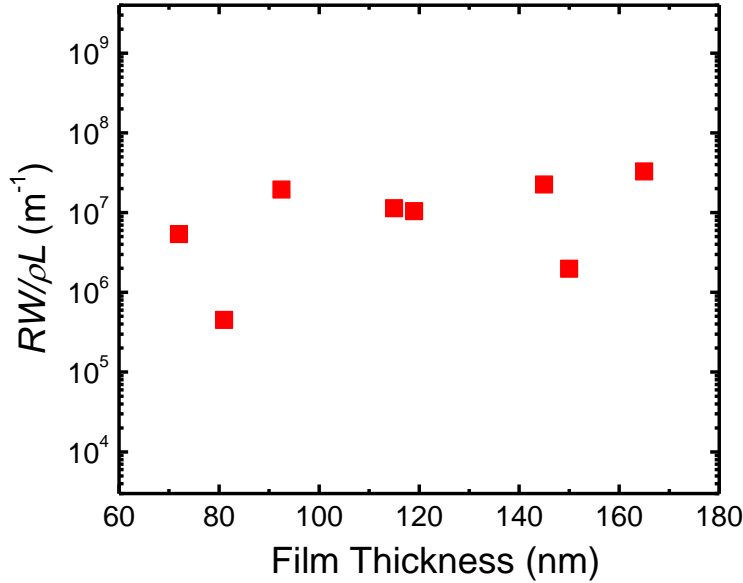

**Figure 5:** Normalized resistance $(R/\rho)(W/L)$ as a function of the film thickness $H$. In the films with the constant resistivity and volume type of electrical conductance in the Ohmic regime one expects $1/H$ dependence of the normalized resistance.

**LOW-FREQUENCY NOISE DATA AND DISCUSSION**

Figure 6 shows a typical noise current spectral density $S_I=S_v[(R_L+R_{ti})/R_L R_{ti}]^2$ for Bi$_2$Se$_3$ devices with Ti/Au contacts as a function of frequency. The noise spectral density $S_I$ of the current fluctuations is approximately inversely proportional to $f$. We determined the dependence by fitting $S_I$ in the form of $1/f^\beta$ to the experimental data. The $\beta$ values varied from 1.09 to 1.13 with an average $\beta=1.11 \pm 0.02$. The absence of the GR bulges on the spectra suggests that there is no a single trap (e.g. at the Bi$_2$Se$_3$/SiO$_2$ interface) that would dominate the noise spectrum like in some semiconductor devices with the thin-film channels.

The low-frequency noise varied with the current as $I^\alpha$, where $\alpha=1.82\pm0.03$. In the Ohmic conduction regime in conventional bulk metals and semiconductors, $S_I$ normally is proportional to $I^2$ (e.g. $\alpha= 2$) [22-24]. The extracted $\alpha$ for the $I^\alpha$ dependence in our devices is close to quadratic but not exactly 2. As we have shown in the previous section, the scaling of the normalized resistance in our samples suggests dominant surface contribution to the electron transport (see Figure 5). The mixed surface-volume type of electrical conduction in Bi$_2$Se$_3$ films may lead to the extracted deviation of $\alpha$ from 2, characteristic for conventional bulk conductors. The measurements of the current fluctuations and noise spectral density in topological insulators can potentially provide additional metrics to be used as indicators for achieving the Dirac cone surface transport regime.

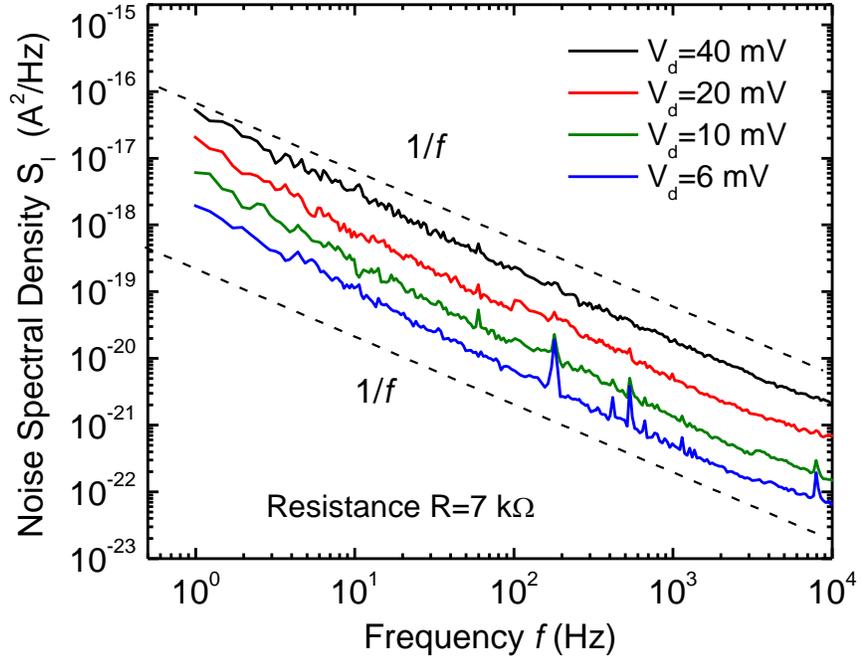

**Figure 6:** Noise power spectral density as a function of frequency for a typical $Bi_2Se_3$ thin-film conductor for bias voltages ranging from 6 mV to 40 mV. The dotted line indicates the 1/$f$ spectrum.

We repeated the noise measurement for a number of devices with various channel sizes characterized by different DC resistances and introduced the amplitude of the current noise as [22, 25] $S_I=AI^2/f^\beta$. Here $A$ is the relative noise amplitude, which is a parameter convenient for comparing 1/$f$ noise levels in various materials and devices. This expression describes well the 1/$f$ noise in all examined devices. While analyzing noise in new material systems it is useful to plot the normalized noise power density as a function of resistance at a fixed frequency [25]. In

Figure 7 we show the normalized noise power ($S_I/I^2$) as a function of the device resistance at $f$=1 Hz. The $Bi_2Se_3$ devices with a higher resistance had the higher normalized noise spectral density. The dash line shown in Figure 7 is the least-squares power-law fit, which yields the following expression for the relative noise amplitude $A$ as a function of the resistance: $A=3\times10^{-10}R^{-0.9}$, where $R$ is the resistance. This expression allows one to estimate an approximate level of $1/f$ noise in $Bi_2Se_3$ thin films for the resistance values spanning over three orders of magnitude. For $Bi_2Se_3$ films we extracted $A$ values in the range from $\sim 10^{-7}$ to $\sim 10^{-5}$.

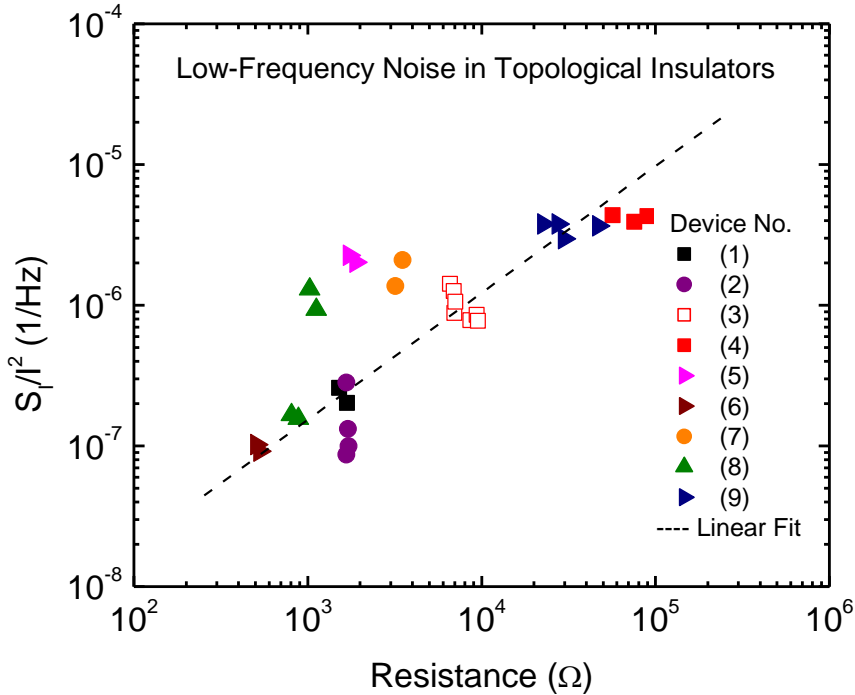

**Figure 7:** Relative noise amplitude $S_I/I^2$ ($f=1Hz$) for several $Bi_2Se_3$ thin-film topological insulators as a function of the device electrical resistance.

Topological insulators and graphene are very different materials in terms of their electrical, mechanical and thermal properties. They do have one feature in common, which allows one to classify them as "Dirac materials": the electron band structure, which reveals a liner dispersion relation within the Dirac cone region. In case of graphene, the electron transport is mostly within the Dirac cone while in topological insulators one usually deals with the mixture of the surface, i.e. Dirac cone, and volume, i.e. parabolic, conduction. It is interesting to note that the low-frequency noise in the examined films made of $Bi_2Se_3$ topological insulators is higher than that in graphene [26-27]. However, the noise spectral density follows the similar dependence on the resistance. From the resistance dependence on the film thickness, we established that in our samples the surface contribution to the electron transport is significant. This means that the measured 1/*f* noise has components originating from both the surface and volume. One can expect that achieving the pure surface conduction regime in topological insulator films would lead to the noise reduction. Moreover, the suppression of at least some electron scattering mechanisms, predicted by the theory [1-8], may results in further noise reduction due to, at least partial, quenching of the mobility fluctuations.

The dimensionless empirical Hooge parameter, $α_H$, [28] defined through the expression $α_H =(S_I/I^2)fN$ (where $N$ is the total number of carriers in the channel) has been conventionally used for comparison of the low-frequency noise level in different materials. The total number of carriers in the channel was estimated as $N=L^2/Rqμ$ within the homogeneous channel approximation (where $L$ is the separation between the contacts, $q$ is the elemental charge and $μ$ is the carrier drift mobility in a given material). The Hooge parameter varies over a wide range of values from the very low, below $\sim 10^{-8}$, to very high, above 10. For example, the value of $α_H$ for *p*-GaN resistor reported in Ref. [29] ranged from 1 to 150. It was found to be between ~0.04 to 0.3 for GaN films with different thickness [30]. The improvements in the GaN film quality (e.g. substantially smaller defect concentration) and increased electron carrier density in 2D channels resulted in much smaller $α_H$ values of $10^{-3} – 10^{-4}$ in GaN heterostructure field-effect transistors [31]. The Hooge parameter $\sim 10^{-3}$ was reported for the long metal resistors or p-n junctions, $\sim 10^{-5}$ for Si complementary metal-oxide-semiconductor devices and $\sim 10^{-8}$ for the short-channel GaAs field-effect transistors and bipolar junction transistors depending on the quality of the materials and devices [31]. Generally a lower value of $α_H$ is expected for the higher level of structural perfection of the materials. At the same time, the Hooge parameter is also affected by the specifics the noise mechanisms and carrier transport.

Although the Hooge parameter for new materials systems has to be taken with reservations, particularly if noise originates on the surface, we calculated it for the films made from topological insulators. To obtain a rough estimate, we took the carrier mobility in $Bi_2Se_3$ as ~180 cm$^2$V$^{-1}$s$^{-1}$ [20] and used the resistance $R$=527 Ω from our DC measurement for the sample with $L$=1.78 μm. The Hooge parameter estimated from the conventional formulas for the homogeneous channel was found to be ~0.2. If a higher mobility of ~500 cm$^2$/Vs [32] is assumed for the Bi$_2$Se$_3$ conduction channel, the Hooge parameter reduces to ~0.07. The upper bound value of 0.2 is higher than that for common semiconductors or metals. At the same time, as the examples of GaN show, this value is not too unusual. The calculation of the total number of carriers using the homogeneous channel expression for Bi$_2$Se$_3$ thin film is not well justified. The "mixed" transport regime, characterized by the conduction through both the volume and surface states, calls for more rigorous procedure for calculation $\alpha_H$. This is not possible at the moment, and requires further theoretical developments in the field of topological insulators.

The reported measurements of the low-frequency current fluctuations in thin films made of topological insulators constitute an additional materials characterization approach for this class of solids. The knowledge of the noise spectral density is essential for any proposed device applications of topological insulators. Whether topological insulators are used in the low-power electronics, memory or quantum computing – the 1/$f$ noise up-conversion due to unavoidable circuit non-linearity will limit the system performance. Based on our results, we also argue that topological insulators may eventually become the current conductors with extremely low noise level. The latter can be achieved if the current goes completely through the surface states protected from scattering by the time-reversal symmetry. In this case, the mobility fluctuation mechanism of the 1/$f$ noise can become suppressed [31]. Determining the relative contributions of different mechanisms, e.g. mobility fluctuations vs. carrier number fluctuations, to the noise characteristics of topological insulators, requires a separate systematic study. It can only be accomplished if one gains a better control of the material quality and position of the Fermi level.

**CONCLUSIONS**

We investigated the low-frequency noise in thin films made of bismuth selenide topological insulator. The thin films were produced by the mechanical exfoliation and used as current conducting channels in the four- and two-contact devices. The thickness of the films was relatively large to avoid hybridization of the top and bottom electron surface states. By investigating the resistance thickness dependence we established that the surface transport was dominant in our exfoliated samples. From the noise measurements it was established that the low-frequency noise spectrum is close to the pure 1/*f*. No generation-recombination peaks were observed in any of the examined devices. The relative noise amplitude $S_I/I^2$ for the examined set of topological insulator films was increasing from ~$5\times10^{-8}$ to $5\times10^{-6}$ (1/Hz) as the resistance of the devices changed from ~$10^3$ to $10^5$ Ω. The measured noise characteristics were attributed to the mixed volume – surface transport regime. The obtained results are important for understanding carrier transport in topological insulators and for their proposed device applications. The knowledge of the noise spectral density and its frequency dependence is also required for design of any practical device made from topological insulator materials.

**MATERIALS AND METHODS**

The thin films of $Bi_2Se_3$ were mechanically exfoliated following the "graphene-like" procedure [33]. The exfoliated films with the lateral dimensions of ~10 μm were transferred to the degenerately doped p-type Si (100) wafers covered with 300 nm of $SiO_2$. Prior to the transfer we cleaned the substrate in acetone and isopropyl alcohol (IPA) solution. The relatively thin flakes were identified using optical microscope and SEM. The crystalinity of the films was check with the micro-Raman spectroscopy (Renisaw, InVia) under the 488-nm excitation laser light. The description of the Raman spectroscopic instrumentation and measurement procedures were reported by us elsewhere [34-35]. The laser excitation power was carefully chosen to avoid any damage to the sample surface while maintaining the acceptable single-to-noise ratio in the measured optical spectra [16-18]. We observed all Raman peaks ($A^1_g$, $E_g$, $A^2_g$) characteristic for crystalline $Bi_2Se_3$. The electron beam lithography (LEO 1550) system was utilized to define the four metal contacts for electrical measurements. The contacts of 20 nm Ti and 180 nm Au were sequentially deposited on the flake surface through the standard electron beam evaporation and lift-off processes. The topological device fabrication and measurements were carried out within a week time. All current – voltage and noise measurements were performed at room temperature under ambient conditions following the standard measurement protocols [36]. The same exfoliation and fabrication techniques and electrical noise measurement procedures can be extended to other materials of the $Bi_2Te_3$ family [16-18, 37].

*Acknowledgements*

The work at UCR was supported, in part, by DARPA – SRC through the FCRP Center on Functional Engineered Nano Architectonics (FENA). The work at RPI was supported, in part, by the NSF Smart Lighting Engineering Research Center and by the NSF I/UCRC "CONNECTION ONE." AAB acknowledges useful discussions on topological insulators with Schoucheng Zhang and Kang L. Wang.